\documentclass{esannV2}

\usepackage{xcolor}
\pagecolor{white}

\usepackage[dvips]{graphicx}

\usepackage[font=scriptsize]{subcaption}
\usepackage[utf8]{inputenc}
\usepackage{amssymb,amsmath,array}
\usepackage{nccmath}

\DeclareMathOperator{\EX}{\mathbb{E}}
\def\G{\mathcal{G}}

\def\V{\mathcal{V}}

\usepackage{booktabs}
\usepackage[font=footnotesize]{caption}

\begin{document}

\title{High Accuracy and Low Regret for User-Cold-Start Using Latent Bandits}
\author{
David Young, Douglas Leith 
\vspace{.3cm}\\
School of Computer Science and Statistics, Trinity College Dublin, Ireland
}
%
%
%
%

\maketitle

\begin{abstract}
We develop a novel latent-bandit algorithm for tackling the cold-start problem for new users joining a recommender system.  This new algorithm significantly outperforms the state of the art, simultaneously achieving both higher accuracy and lower regret. 
\end{abstract}

\section{Introduction}
In a recommender system, when a new user joins the system it initially has no knowledge of the preferences of the user and so would like to quickly learn these.
The recommender system therefore initially starts in an ``exploration'' phase where the first few items that it asks the new user to rate are chosen with the aim of discovering the user's preferences.   We focus on the simplest setup where a user explicitly rates items presented to them, e.g. on a 1-5 scale, and the aim of the recommender system is to predict other items that the user may like.

\begin{figure}
\centering
\begin{subfigure}[t]{0.25\textwidth}
\includegraphics[width=\textwidth]{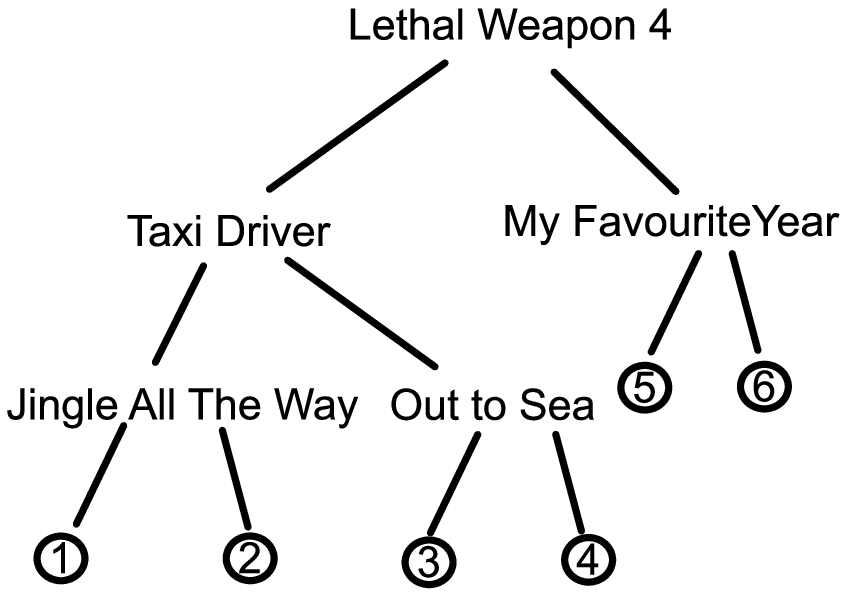}
\caption{\scriptsize}
\end{subfigure}
 \hspace{5mm}
\begin{subfigure}[t]{0.25\textwidth}
\includegraphics[width=\textwidth]{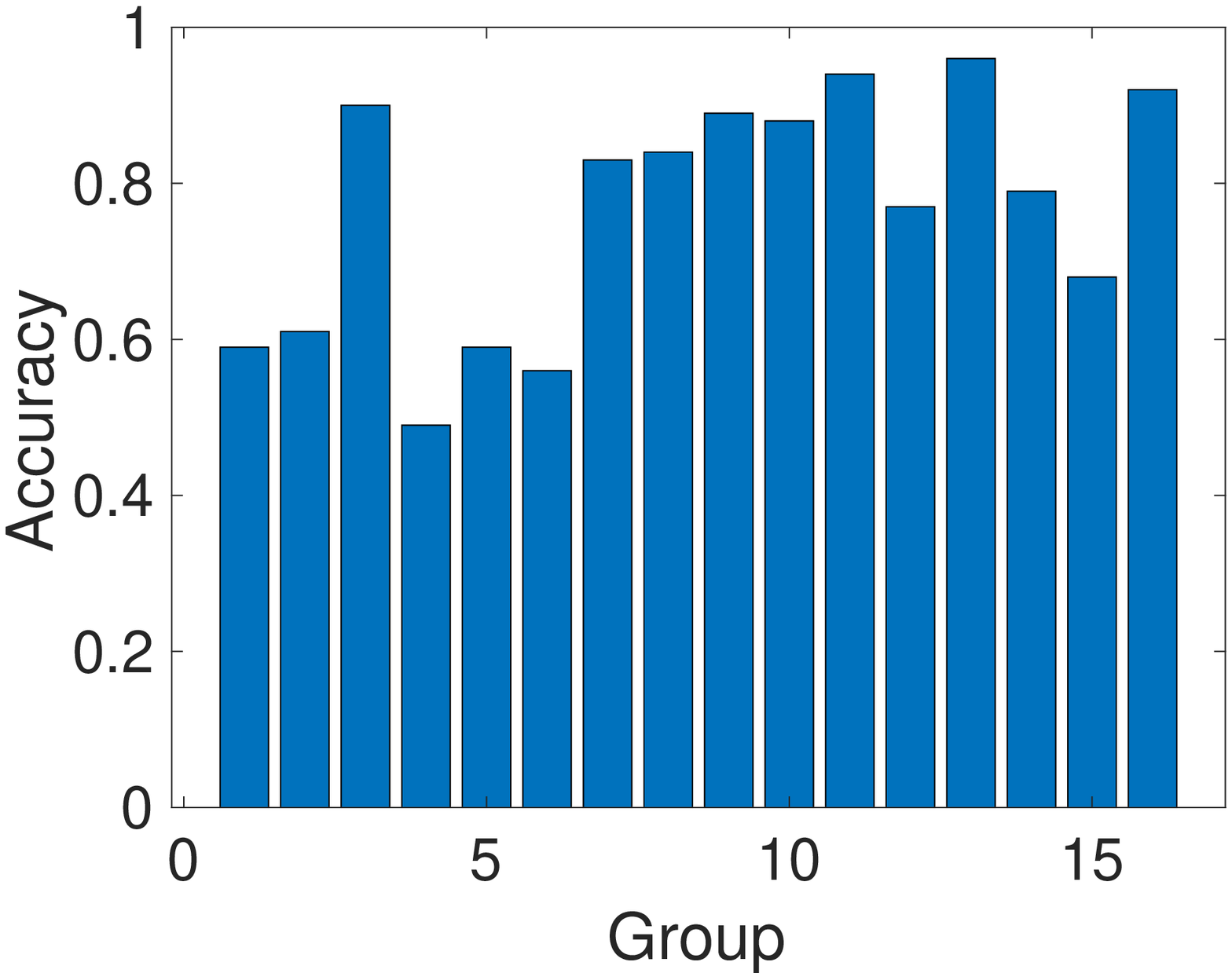}
\caption{\scriptsize}
\end{subfigure}
\vspace{-0.25cm}
\caption{\scriptsize (a) Illustrating a movie recommender decision-tree (b) Decision-tree accuracy for Netflix data (16 groups).}\label{fig:one}
\vspace{-0.3cm}
\end{figure}

One common approach to this new user cold-start task is to take ratings already collected from a population of users, use these to cluster users into groups and then train a decision-tree to learn a mapping from item ratings to the user group, see for example Figure \ref{fig:one}(a).   When a new user joins the system this decision-tree is used to decide which items the user is initially asked to rate and in this way the group to which the user belongs is initially estimated.   Once the group is estimated, the system recommends items liked by members of that group e.g. using matrix factorisation or another collaborative filtering approach.  

However, typically users clustered in the same group do not give identical ratings to an item.  Rather there is a spread of ratings, and this intra-cluster variability between users can be thought of as adding noise to the ratings.  Decision-trees are vulnerable to such noise in the new users rating as an unusual rating for a particular group can send the tree down a wrong path it will never recover from.   For example,  Figure \ref{fig:one}(b) shows the measured decision-tree accuracy for Netflix data clustered into 16 groups (see later for details).  It can be seen that the accuracy is as low as 50-60\% for some groups.

In this paper we improve on this behaviour by developing a new online learning algorithm that maintains and updates a probability distribution for the users group, and then selects the items that a new user is asked to rate so that this distribution converges to correct group as quickly as possible. In this way for users with more noise in their ratings it will simply take the system longer to learn the correct group, instead of possibly concluding the wrong group.    

To develop our new online learning algorithm we view the cold-start task as a latent bandit, i.e. a multi-arm bandit where the distribution of the arms depends on the value of someone unknown latent parameter.  In our setting the arms are the available items, the reward of an arm is the user rating for the corresponding item and the latent variable is the true group of the user.    It is important to stress that ignoring the latent groups and applying standard bandit algorithms to this task leads to poor performance since (i) there are many arms and so learning is slow and (ii) repeated pulls of the same arm tend to be highly correlated.
In contrast, the latent group can take a small number of values, e.g. there might only be 16 or 32 groups, and for each value there is a known distribution of arm rewards.   The latter means that we do not need to pull every arm to gain information about its reward and so the latent group approach allows fast learning even when the number of arms is large.   

Further, in general, some arms tend to be more informative than others for learning the user group.  Hence, pulling only the most informative arms allows us to quickly learn the user group.  Importantly, provided we have a sufficient number of informative arms then fast learning can still be achieved even when each arm is only pulled once i.e. a new user is only asked to rate any given item once. 
We want to select the next arm to pull based on our current estimate of the probability distribution for the users group and with the aim of causing our estimated distribution to quickly concentrate on the true user group.  Note that existing methods for latent bandits require repeated arm pulls and do not take full advantage of the informative arms.

\section{Related Work}

We follow on from the work of \cite{shams2021cluster} in using cluster based bandits to tackle the cold start problem.  
Our latent bandit approach bears most similarity to the work of \cite{hong2020latent}, which develop algorithms based on UCB and Thompson sampling for the same setup. 
 Our algorithm however leverages the latent-bandit in a way their methods do not. In \cite{pmlr-v32-maillard14}, they first apply a UCB style algorithm to the same problem, before relaxing assumptions and proposing algorithms for problems where the reward distributions are unknown.
A survey of active learning cold-start methods can be seen in \cite{article}, and both \cite{10.1145/1540276.1540302} and \cite{10.1145/1935826.1935910} use decision tree based methods in a similar cluster based setup to ours. 

\section{Latent Bandit Algorithm For User Cold-Start}
We have a set $\G$ of user groups, set $\V$ of items.  
Given a new user our task is to quickly learn which group $g\in\G$ they belong to by asking the user to rate items in $\V$, using the fact that the distribution of item ratings varies depending on the user's group.  

\subsection{User Ratings}
We assume the rating of item $v$ by users in group $g$ is normally distributed with mean $\mu (g,v)$ and variance $\sigma^2(g,v)$.   If a user $u$ in group $g$  is asked to rate the same item $v$ multiple times the user gives the same rating i.e. their rating is one value drawn from $N(\mu (g,v),\sigma^2(g,v))$.   
%
%
Let random variable $R(v)$ be the rating of item $v$ and random variable $G(u)$ be the group of the user making the rating.  Then $p(R(v)=r | G(u)=g)= (1/\sqrt{2\pi}\sigma(g,v))e^{-(r-\mu(g,v))^2/2\sigma^2(g,v)}$ and for observed sequence $D_n$ of ratings $r_1,r_2,\dots,r_n$ for items $v_1,v_2,\dots,v_n$ ($D_n$ is a sequence of pairs $(v_i,r_i), i=1,\dots,n$)
it follows that 
$$p(D_n | G(u)=g)= \gamma_n(g)e^{-L_n(g)}$$
where $L_n(g):=\sum_{i=1}^n(r_i-\mu(g,v_i))^2/2\sigma^2(g,v_i)$, $\gamma_n(g) := 1/(2\pi)^{n/2}\times 1/\Pi_{i=1}^{n}\sigma(g,v_i)$.  By Bayes rule
$$ p(G(u)=g | D_n) = \frac{p(D_n | G(u)=g)p(G(u)=g)}{p(D_n)}$$
with $p(D_n) = \sum_{h\in \G}$ $ p(D_n | G=h)p(G(u)=h)$.  Assuming uniform prior $p(G(u)=g)=1/|\G|$ then the probability that a new user belongs to group $g$ given item ratings $D_n$ is
\begin{align}
	p(G(u)=g | D_n) = \frac{p(D_n | G(u)=g)}{\sum_{h\in \G} p(D_n | G(u)=h)}
	=\frac{\gamma_n({g})e^{-L_n ({g})}} {\sum_{h\in \G}\gamma_n(h) e^{-L_n(h)} }\label{eq:zero}
\end{align}


\subsection{Exploration: Knowledge Gained From a New Rating}

Given a new user's ratings $r_1,r_2,\dots,r_n$ for a sequence of items $v_1,v_2,\dots,v_n$, we need to select the next item $v_{n+1}$ to ask the user to rate.   We have a current estimate $P(G(u)=g| D_n), g\in\G$  of the probability distribution of the user group.  Intuitively, a reasonable choice is the item that is most likely to cause this distribution to maximally concentrate on the true useruser group $g^*$.  This means  selecting the item  $v_{n+1}$ which maximises
$$\EX[P(G=g^*|D_n, v_{n+1}, R(v_{n+1}); G = g^*)$$
Since in reality we don't know $g^*$ we want to select the item $v_{n+1}$ that maximises the expected value with respect to $g^*$, i.e.
\begin{align*}
  \EX_{g^*}[\EX[P(G=g^*|D_n, &v_n,R(v_{n+1}); G = g^*)]] \\
  &= \sum_{g \in \G} P(G=g|D_n)\cdot \EX[P(G=g|D_n, v_{n+1},R(v_{n+1});G=g)]
\end{align*}
This expected value cannot be found analytically, but if we take a linear approximation of $P(G=g^*|D_n, v_n,R(v_{n+1}); G = g^*)$ and take our expectation over that we obtain
\begin{equation}\label{eqn:apr}
	\EX[P(G=g^*|D_n, v_n,R(v_{n+1}); G = g^*) ]\approx  P(G=g^*|D_n, v_{n+1}, \mu(g^*, v_{n+1}))
\end{equation}
%
%
Instead of the approximation \eqref{eqn:apr} we could use Monte Carlo simulation to evaluate this expectation but this is considerably slower to calculate and in our tests comparing \eqref{eqn:apr} with the values calculated by Monte Carlo we find that \eqref{eqn:apr} has surprisingly small approximation error, and negligible effect on performance.

\subsection{Exploitation: Future Reward}
Selecting the next item $v_{n+1}$ to maximise \eqref{eqn:apr} prioritises learning about the group that a new user belongs to i.e. exploration.   However, items which accelerate learning may attact a low user rating, and so increase regret.  
We need, therefore, to balance exploration against exploitation i.e. selecting items predicted to have a high user rating.   

However, when considering exploitation it is necessary to take account of the uncertainty in our current estimate of the user's group.  This is because items rated highly by users in one group may not be rated highly by users of another group.  Hence, if we make a mistake in our estimate of the new user's group we may end up suggesting items that return a low rating by the user and so increase regret.

We proceed by defining the discounted future reward.  For a user in group $g$ who has already rated items $V_n=\{v:(v,\cdot)\in D_n\}$ the expected future reward is $\sum_{v\in V\setminus V_n} \mu (g,v)$, assuming they stay in the system and eventually rate all items.  However, its probably more reasonable to assume there is a departure process for users, who tend to only stay in the system for some lifetime.  For simplicity, we will assume the case where the departure process is modelled as an independent event after every recommendation, with constant probability $\beta$ of staying in the system. We could replace this with any step dependent model for the probability of the user still being the system. We then use $\beta$ to discount future rewards. Let $V_{n,g}^*$ be the sequence of items $V\setminus V_n$ sorted in decreasing order of mean rating $\mu (g,v)$.   Assuming the recommender presents items to the user in this order, then the expected discounted future reward is
\begin{align*}
	J_{future}(g):=\sum_{i=1}^{|V_{n,g}^*|} \beta^i\mu (g,v_{i,g}) 
\end{align*}
where $v_{i,g}$ is the $i$'th element in sequence $V_{n,g}^*$ and $0 \le \beta\le 1$ is our discount factor.  And so the expected discounted future loss of acting as if the user is in group $h$ when its actually in group $g$:
\begin{align*}
	J_{future loss}(g,h):=\sum_{i=1}^{|V_{n,g}^*|} \beta^i|\mu (g,v_{i,g})-\mu(g,v_{i,h})| 
\end{align*}
where $v_{i,h}$ is the $i$'th element in sequence $V_{n,h}^*$. Since we have estimates $P(G(u)=g| D_n), g\in\G$  of the probability distribution of the user group, we can calculate the expected discounted future regret of acting as if the user is in group $h$, at any step $n$ is :
\begin{align*}
	J_{future regret}(h):=\sum_{g\in G}P(G=g|D_n) \sum_{i=1}^{|V_{n,g}^*|} \beta^i|\mu (g,v_{i,g})-\mu(g,v_{i,h})|  
\end{align*}
This value is exactly the expected opportunity cost of exploiting instead of learning more information about which group the user is in. 


\subsection{Balancing Exploration \& Exploitation: New Algorithm}


We balance exploration and exploitation by selecting the next item $v_{n+1} $ that maximises
\begin{equation*}\medmath{
    v_{n+1}  \in \arg \max_{v \in V\setminus V_n}\sum_{g\in\G}P(G = g|D_n) \cdot P(G=g|D_n, v, \mu(g, v))\cdot [J_{future regret}(g)^2 + \mu(g,v)]
  }
 \end{equation*}

When the opportunity cost $J_{future regret}(g)$ of incorrectly estimating the user group is large, the short term reward $\mu(g,v)$ is effectively ignored and the next item is selected primarily to minimise $J_{future regret}(g)$ i.e. to maximally increase the accuracy of our estimate of the user's group.  Conversely, when $J_{future regret}(g)$ is small, the next item is selected to maximise the short term reward $\mu(g,v)$ i.e. picking an item likely to be rated highly by the new user given our current best estimate of the user's group.



\begin{figure}
	\centering
\begin{subfigure}[t]{0.35\textwidth}
		\includegraphics[width=\textwidth]{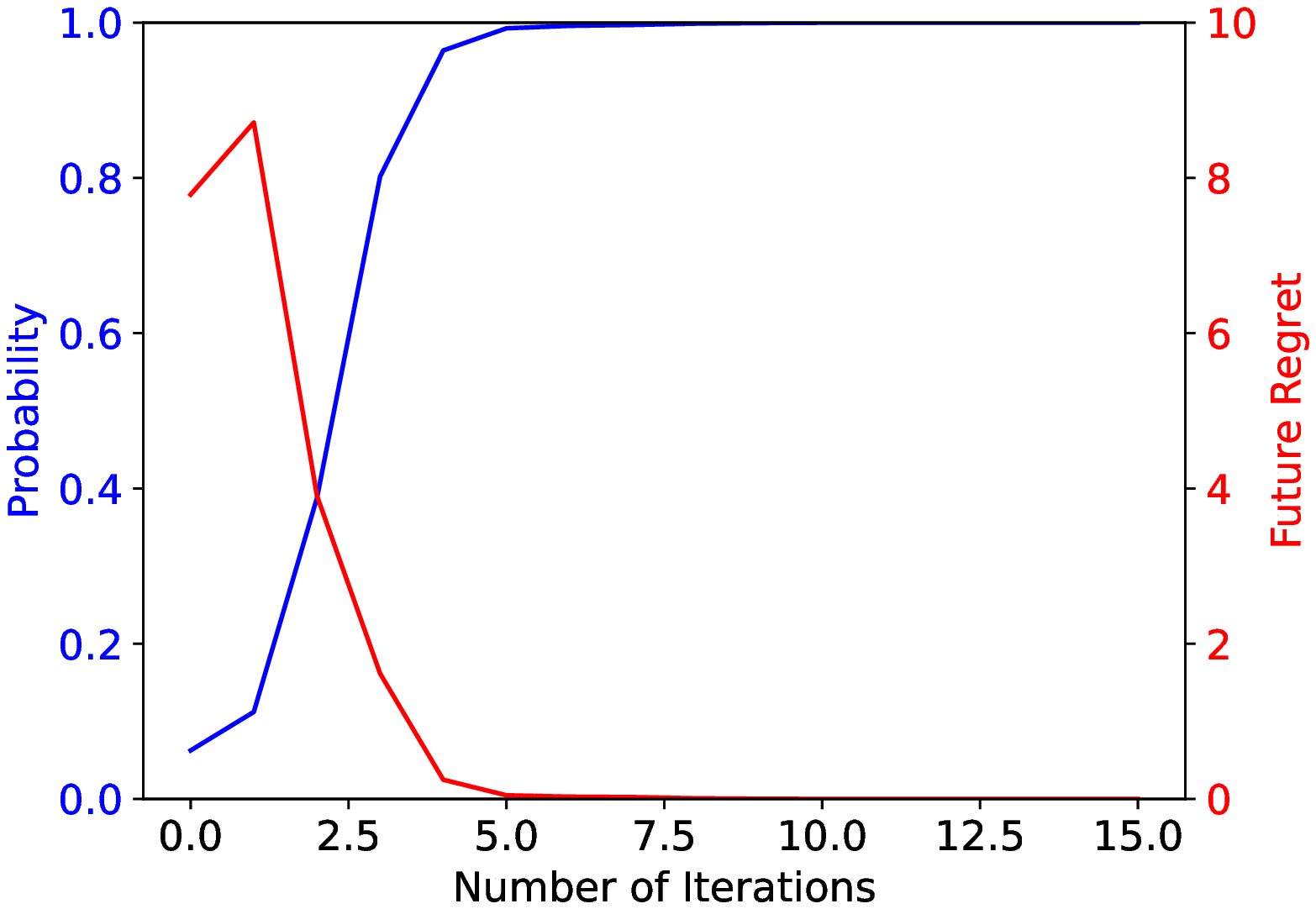}
		\caption{ }
\end{subfigure}
\begin{subfigure}[t]{0.35\textwidth}
		\includegraphics[width=\textwidth]{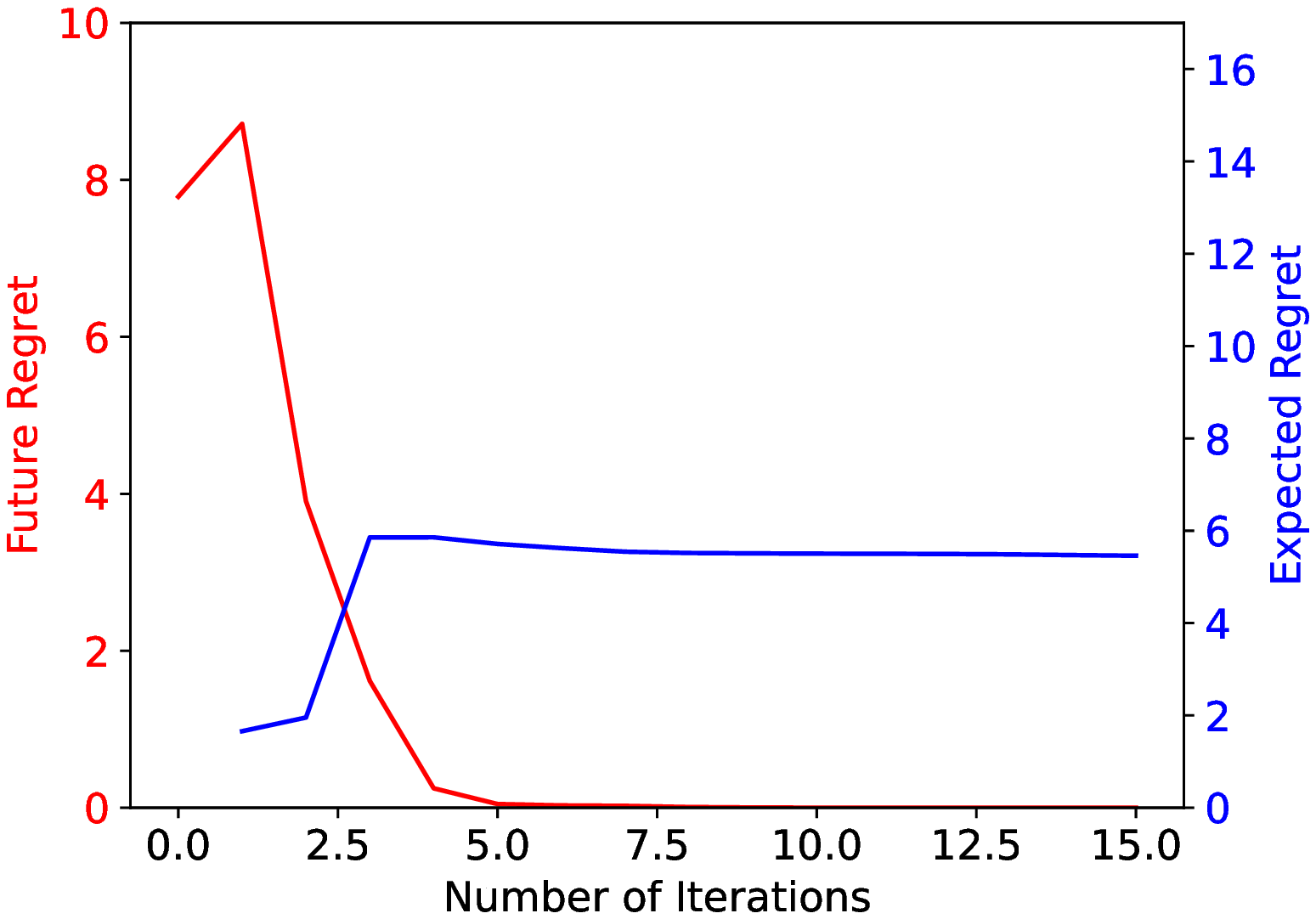}
		\caption{}
\end{subfigure}
\vspace{-0.25cm}
\caption{\scriptsize	Measurements illustrating the transition from exploration to exploitation.  Netflix dataset with 16 groups and a new user from group ten. }\label{fig:net16prob_fr}
	\vspace{-0.3cm}

\end{figure}

Figure \ref{fig:net16prob_fr}(a) shows measurements illustrating the transition from exploration to exploitation for the Netflix dataset with 16 groups and a new user from group ten.   It can be seen that the future regret for exploiting as if we are in group ten is initially high, as the probability for being in any particular group is low.   The probability and future regret then rise together, indicating the user has given a rating that also increases the probability of being in groups for which the top rated items are very different. As the probability of being in group ten continues to rise sharply, the future regret of exploiting as if we are group ten drops sharply, as expected.  Figure \ref{fig:net16prob_fr}(b) shows the relation between the future regret value and the actual expected regret incurred by the system. It can be seen that the regret incurred rises steeply initially while the future regret is high, as the system focuses on learning more about the users group. Then as the future regret of exploiting as if we are in group ten becomes very small, the system starts to do just that, and the expected regret no longer grows. 

\section{Performance Evaluation}
\qquad \emph{Datasets.}  We evaluate the performance of our latent bandit algorithm on the standard Netflix dataset 
, the Jester dataset 
and the Goodreads10K dataset. 

\emph{Clustering Users.} We use training data to cluster users into groups and estimate the mean $\mu(g,v)$ and variance $\sigma(g,v)^2$ of the ratings by each group $g$ for item $v$.  

\emph{Baseline Algorithms.} We compare the performance of the latent bandit algorithm (LBA) against (i) an optimised CART decision tree and (ii) the cluster-based bandit (CBB) algorithm of~\cite{shams2021cluster}.   These are strong baselines, with good performance for cold-start active learning.   Decision-trees are often considered for use in cold-start while the recently proposed CBB-algorithm offers state of the art performance.

\emph{Modelling New Users.} We generate the item ratings of a new user from group $g$ by making a single draw from the multivariate Gaussian distribution with mean $\mu(g,v)$ and variance  $\sigma(g,v)^2$ for each item equal to that estimated from the training data.  This has the advantage that we can easily generate large numbers of new users in a clean, reproducible manner.  

\emph{Performance Metrics.}  We report the accuracy with which the group of a new user is estimated, i.e. the fraction of times the correct group is estimated, and the expected regret.  Statistics are calculated over 1000 new users per group.


\begin{figure}
	\centering
	\begin{subfigure}[t]{0.35\textwidth}
		\includegraphics[width=\textwidth]{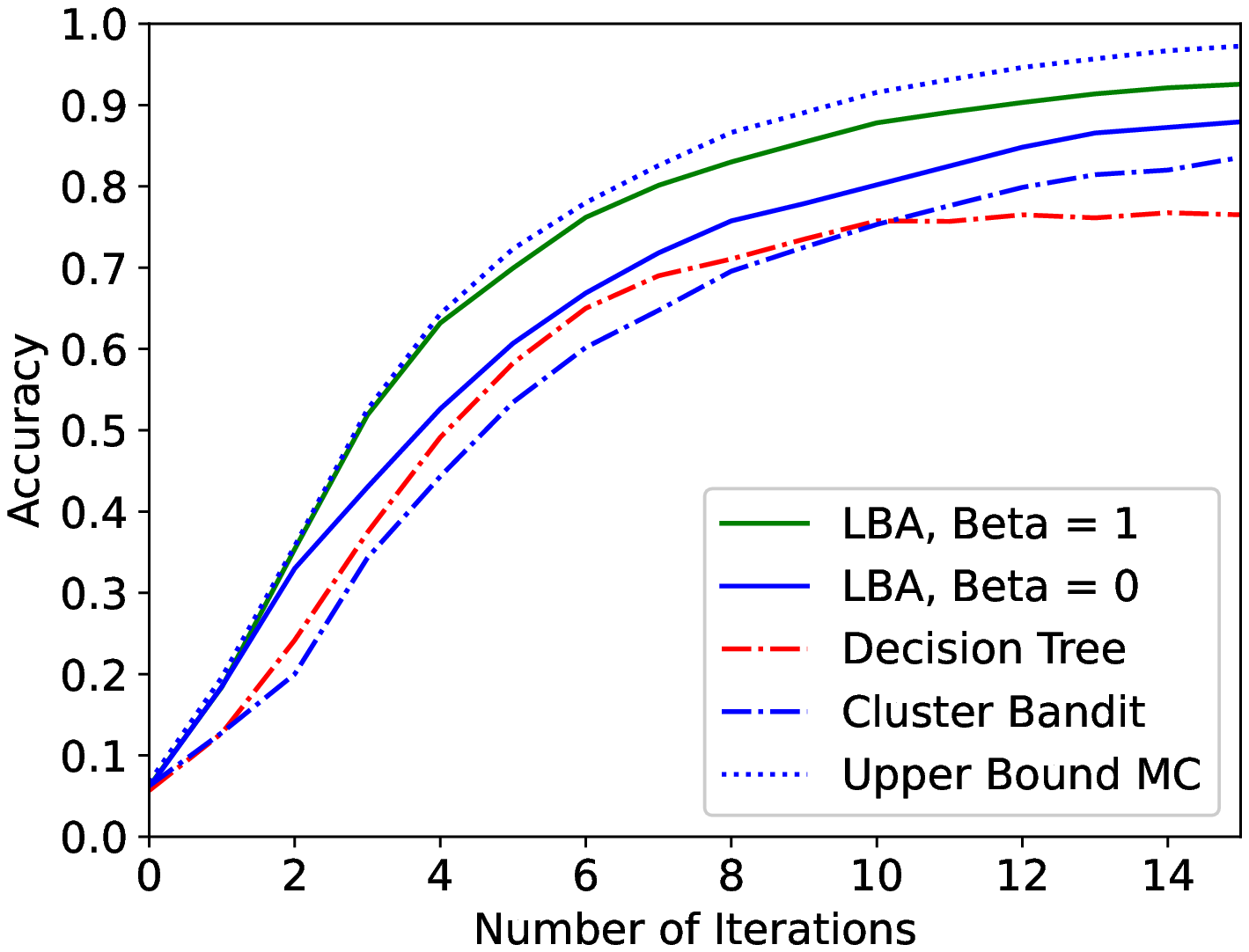}

	\end{subfigure}
	\begin{subfigure}[t]{0.35\textwidth}
		\includegraphics[width=\textwidth]{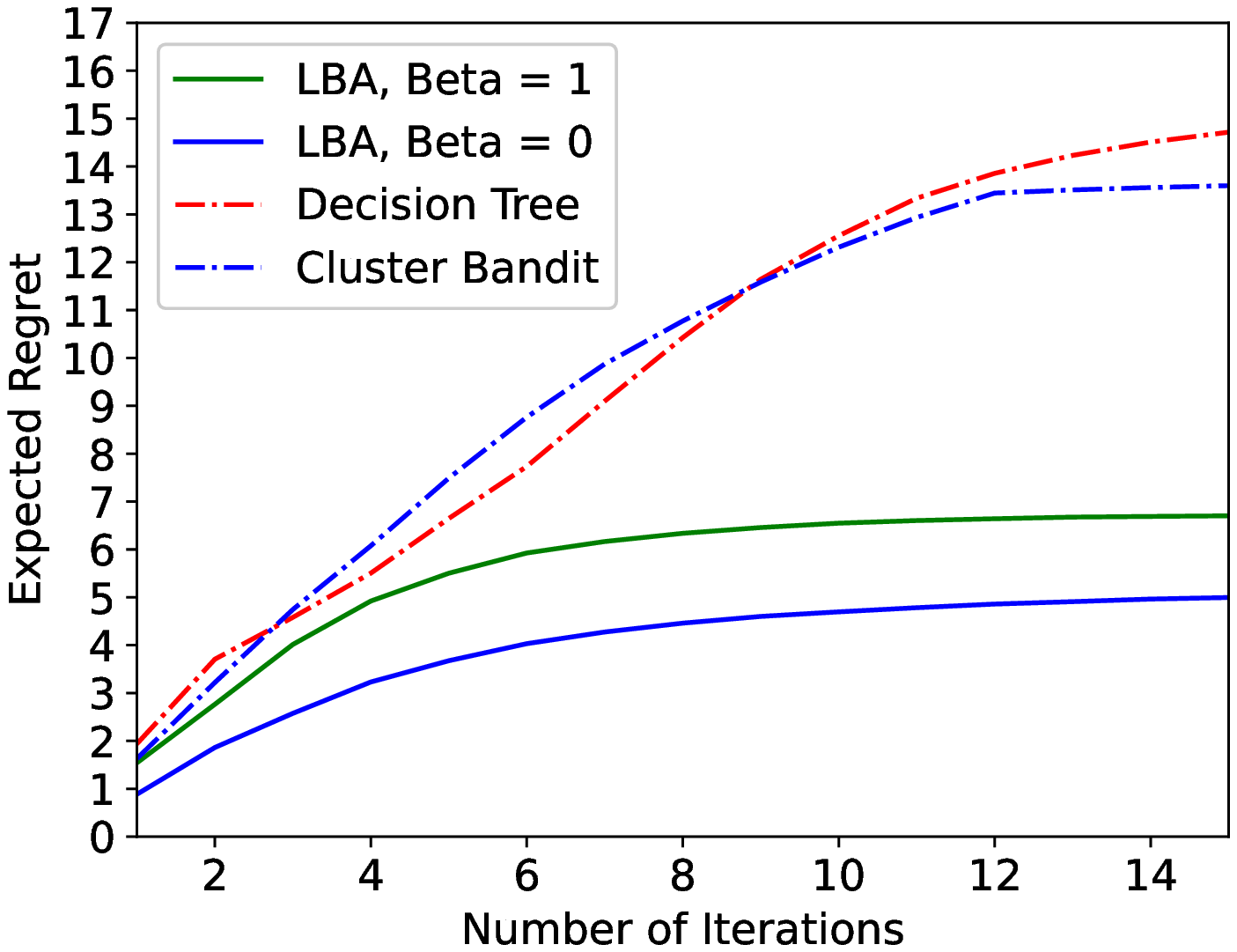}
	\end{subfigure}
	\vspace{-0.4cm}
	\caption{\scriptsize Accuracy and expected regret averaged over all groups for Netflix dataset with 16 clusters. The upper bound we plot is  the accuracy of our algorithm focused exclusively on learning, with the expectation calculated using MC. }\label{fig:net16acc}
\vspace{-0.25cm}
\end{figure}
\begin{figure}
	\centering
	\begin{minipage}{0.35\textwidth}
		\centering
		\includegraphics[width=\textwidth]{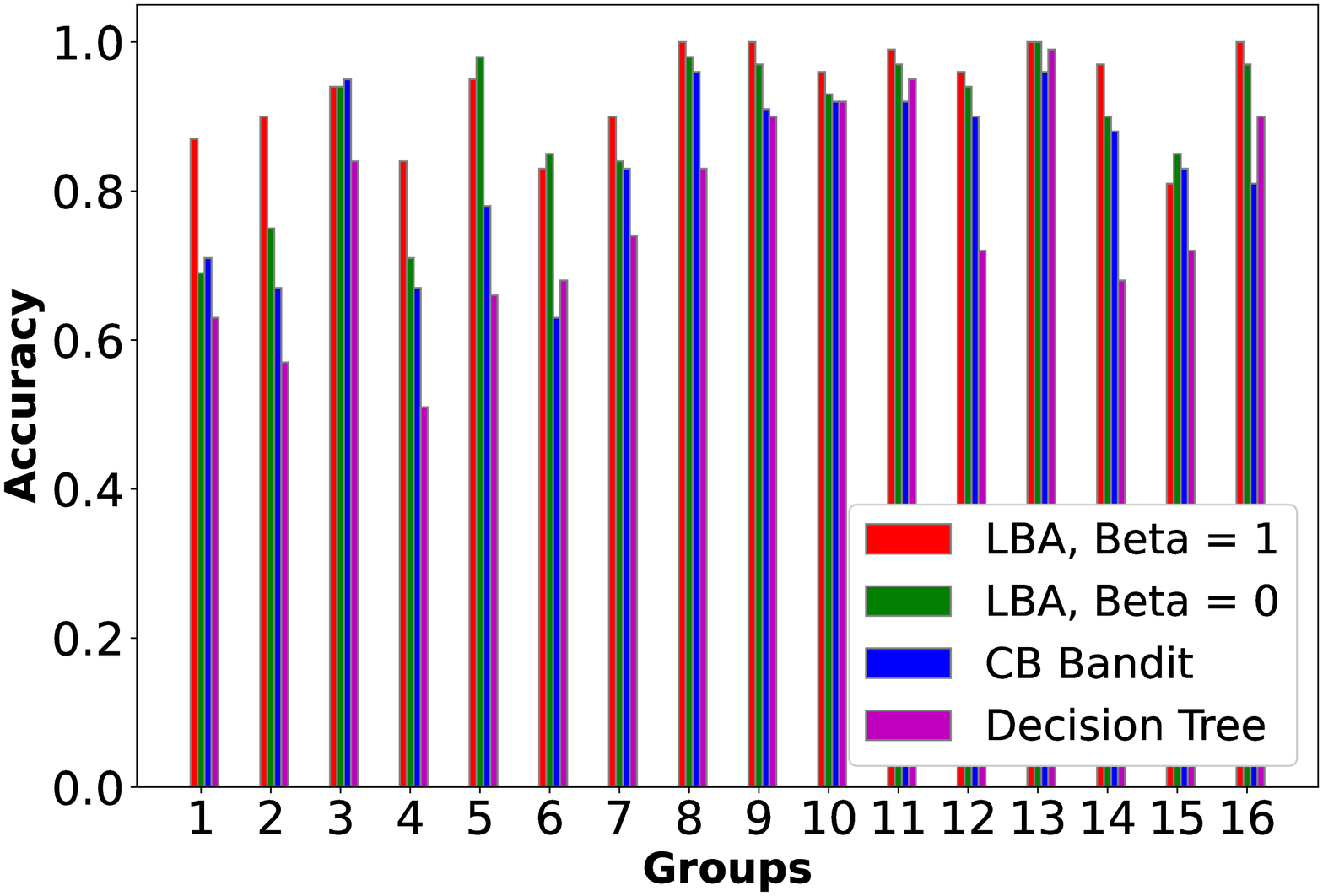}
	\caption{\scriptsize Accuracy after 15 iterations vs user group.  Netflix dataset with 16 clusters.}
		\label{fig:net16_acc_bar}
	\end{minipage} \quad
	\begin{minipage}{0.6\textwidth}
		\centering
		\scriptsize
		\scalebox{0.58}{\begin{tabular}{lcccccccl}\toprule
				& \multicolumn{4}{c}{Accuracy} & \multicolumn{4}{c}{Regret}
				\\\cmidrule(lr){2-5}\cmidrule(lr){6-9}
				& \multicolumn{4}{c}{No. of Groups} & \multicolumn{4}{c}{No. of Groups}
				\\\cmidrule(lr){2-5}\cmidrule(lr){6-9}
				Dataset/Algo& 4             & 8       &  16                & 32                 & 4               & 8               &   16             & 32\\\midrule
				Netflix/DT    & 0.92       & 0.9    & 0.77               & 0.55              & 9.6        & 12.1             &  14.8          & 20.2\\
				Netflix/CB    & 0.96       & 0.93   & 0.89             &  0.74             & 5.1         & 9.8               & 13.9               & 16.6\\
				Netflix/LBA,0& 1             & 0.98   & 0.94              &   0.83        &\textbf{2.1} & \textbf{4.3}& \textbf{5.1}& \textbf{5.5}\\
				Netflix/LBA,1 & \textbf{1} &\textbf{1} &\textbf{0.98}& \textbf{0.91}&   4.1   & 7.2            & 9.7             &  12.4\\\midrule 
				Books/DT & 0.92       & 0.7         & 0.61               & 0.4                         & 3        & 9.5             &  8.4          & 8.4\\
				Books/CB& 0.97       & 0.88        & 0.84               & 0.66                       & 5.7        & 8.4             &  9          & 6.1\\
				Books/LBA,0& 0.99      & 0.9       & 0.9                  &0.77                    &\textbf{1.1} & \textbf{3}& \textbf{3.6}& \textbf{3}\\
				Books/LBA,1& \textbf{1} &\textbf{0.9} &\textbf{0.92}& \textbf{0.8}  &   1.1        & 3.5           & 4.7             &  5.2\\\midrule 
				Jester/DT & 0.7                          & 0.54     & 0.41               & 0.34              & 21.3        & 27.9             &  35.4          & 43.3\\
				Jester/CB& 0.91                         & 0.82       & 0.76               & 0.68             & 15.9        & 22.1             &  33          & 53.8\\
				Jester/LBA,0& 0.93                   & 0.85      & 0.78              &   0.69        &\textbf{1.7} & \textbf{6.1}& \textbf{9.7}& \textbf{13.8}\\
				Jester/LBA,1& \textbf{0.94}&\textbf{0.86} &\textbf{0.85}& \textbf{0.85}&   1.8        & 8.9           & 23.4             &  40\\\bottomrule
				
		\end{tabular}}
		\caption{\scriptsize Performance after 25 iterations, averaged over all groups.   }
		\label{table:acc_reg}
	\end{minipage} 
	\vspace{-0.3cm}
\end{figure}

Figure \ref{fig:net16acc} shows typical measurements of the evolution of accuracy vs \#items rated by a new user.  It can be seen that the accuracy grows over time and that the performance of the new latent-bandit algorithm dominates that of the decision-tree and cluster-based bandit.  Data is show for future regret discount factor $\beta$ of 0 and 1, when $\beta=1$ the latent-bandit focuses on exploration until there is almost no more possible gain from it, and it can be seen that, as expected, the learning rate is somewhat faster than when $\beta$ is smaller, but this is balanced against increased regret (which nevertheless remains uniformly lower than that of the  decision-tree and cluster-based bandit). The dotted line in left-hand plot shows the performance of the latent-bandit focused exclusively on exploration, with the expectation calculated using Monte Carlo. It can be seen that with $\beta=1$ the latent-bandit performance is close to this upper bound, indicating little scope for improvement.  Figure \ref{fig:net16_acc_bar} breaks this accuracy data down by group, observe that the variability in accuracy amongst the groups is significantly reduced by the latent bandit algorithm.  Figure \ref{table:acc_reg} shows summary data for the Netflix, Jester and Goodreads datasets and for 4 to 32 user groups.   It can be seen that the uniformly outperforms the state of the art, simultaneously achieving both higher accuracy and lower regret.


\section{Conclusion}
We present a novel latent-bandit algorithm for tackling the cold-start problem for new users joining a recommender system.  This new algorithm uniformly outperforms the state of the art, simultaneously achieving both higher accuracy and lower regret. 


\begin{scriptsize}




\bibliographystyle{unsrt}
\bibliography{bib}

\end{scriptsize}


\end{document}